\documentclass[fleqn,twoside]{article}
\usepackage{espcrc2}

\usepackage{graphicx}
\usepackage[figuresright]{rotating}

\newcommand{\AmS}{{\protect\the\textfont2
  A\kern-.1667em\lower.5ex\hbox{M}\kern-.125emS}}

\title{Automatic one-loop calculation of MSSM processes with GRACE}

\author{J. Fujimoto\address[KEK]{KEK, Oho, Tsukuba, Ibaraki 305-0801 Japan},
        T. Ishikawa\addressmark[KEK],
        M. Jimbo\address[TMC]{Tokyo Management College, Ichikawa,
        Chiba 272-0001, Japan},
        T. Kon\address[SU]{Seikei University, Musashino, Tokyo 180-8633, Japan}
        and
        M. Kuroda\address[MGU]{Meiji Gakuin University, Totsuka,
        Yokohama 244-8539, Japan}}

\begin{document}

\begin{abstract}
We have developed the system for the automatic computation of cross sections,
{\tt GRACE/SUSY}~, including the one-loop calculations for processes of
the minimal supersymmetric extension of the the standard model.
For an application,
we investigate the process $e^+ e^- \rightarrow Z^0 h^0$~.
\vspace{1pc}
\end{abstract}

\maketitle

\section{Introduction}

Supersymmetry (SUSY) between bosons and fermions at the unification-energy
scale is one of the most promising hypothesis, which is expecteted to resolve
the remaining problems in the standard model (SM).  In particular, the minimal
supersymmetric extension of the SM (MSSM) \cite{theor} has been
extensively studied in the last decade due to the simplicity.

For more than ten years, we have been developing the system of the automatic
computation of the high energy physics processes.  The system for the
computation of the SM, {\tt GRACE}, has been published in \cite{grace}.

In including the interactions of SUSY particles in the {\tt GRACE} system,
we have made several modifications and the expansion of the system
\cite{gs,maj}.  As the first outcome from {\tt GRACE/SUSY}, we have published
a package of event-generator, {\tt SUSY23}, which contains 23 specific SUSY
processes for $e^+e^-\to$ 2-body and 3-body \cite{s23}.  At this stage,
the model definition files are based on the Hikasa's Manual \cite{kh}.

Recently, we have constructed the complete lagrangian of the MSSM \cite{mk}
using
the European convention: namely, the positive chargino is called a particle and
the ranges of $\mu$ and $\tan\beta$ are defined as $0\le \tan\beta \le 1$ and
$-\infty\le \mu \le+\infty$~.  Thus we have published the new version of
{\tt GRACE/SUSY} ({\tt GRACE}~{\tt v2.2.0}) \cite{gsp}, which is available
from {\tt http://minami-home.kek.jp/}~.

In the world, there exist several other groups independently developing the
systems of the automatic computation in the SM with different methods
\cite{comp,Den,Mad,THahn}, and also developing the systems of the automatic
computation in the MSSM, {\tt FeynArts-FormCalc}~\cite{THahns} and
{\tt CompHEP}~\cite{comps}.

In this paper, we present the latest development of the {\tt GRACE/SUSY} system
including the one-loop calculations in the MSSM.

\section{\tt \bf GRACE/SUSY/1LOOP}

\subsection{Renormalization scheme}

In the the MSSM, several particles are mixed states, so there are three kinds of
way of
introducing wavefunction renormalization constants.  We adopt the
renormalization scheme of the MSSM as follows:
\begin{itemize}
\item the gauge-boson sector: the conventional approach \cite{conv}\\
(Renormalization constants of wavefunctions are introduced to unmixed bare
states and mass counterterms are introduced to mixed mass eigenstates.)
\item the Higgs sector: the Dabelstein's approach \cite{Dab}~;
 the chargino sector and the neutralino sector: the Kuroda's approach \cite{MK}
(see also \cite{Fri})\\
(Renormalization constants of wavefunctions are introduced only to unmixed
bare states.)
\item the matter-fermion sector and the sfermion sector: the Kyoto approach
\cite{kyoto}\\
(Renormalization constants of wavefunctions are introduced only to mixed
mass eigenstates.)
\end{itemize}

\subsection{How to check the system}

For the tree-level calculations, we first check the gauge invariance of
amplitudes at a point of the phase space before the integration.  In the
{\tt GRACE} system, the gauge invariance check is automatically carried out
using the covariant gauge and the unitary gauge.  In the SM, we have also
checked
{\tt GRACE} with the non-linear gauge \cite{nlg}.  In the MSSM, we have already
checked
the gauge invariance for
582,102 processes with up to six-external particles within quadruple precision
\cite{gauge}.

For the one-loop calculations, we check the invariance of cross sections
varying three parameters, the UV constant ($C_{\rm UV}$), the fictitious photon
 mass
($\lambda$) and the cutoff energy of the soft photon ($k_{\rm c}$).  As an
example of the invariance checks, the result for the process
${e^+ e^- \rightarrow {\tilde{\chi}_{\rm 2}}^{+} {\tilde{\chi}_{\rm 2}}^{-}}$
at $\sqrt{s}=1900$~GeV is shown in ref.~\cite{Our}, using the
same input parameters as in ref.~\cite{diaz}.

\begin{table*}[htb]
\label{table:eq1}
\renewcommand{\tabcolsep}{1.2pc} % enlarge column spacing
\begin{tabular}{|r|r|r|r|r|r|r|}
\hline
$\tan\beta$ & \multicolumn{1}{c|}{$m_{h^0}$}
            & \multicolumn{2}{c|}{$M_{h^0}$}
            & \multicolumn{1}{c|}{$m_{H^0}$}
            & \multicolumn{2}{c|}{$M_{H^0}$} \\
\hline
            & \multicolumn{1}{c|}{tree} & 1-loop & Dabelstein
            & \multicolumn{1}{c|}{tree} & 1-loop & Dabelstein \\
\hline
  0.5  &  53.11506 &  93.1  &  93.2 & 309.0208 & 352.4 & 352.1 \\
  2.0  &  53.11506 &  83.8  &  84.0 & 309.0208 & 312.3 & 312.3 \\
  5.0  &  83.55589 & 106.0  & 106.2 & 302.2143 & 302.9 & 303.0 \\
 10.0  &  89.20395 & 110.5  & 110.8 & 300.5955 & 300.8 & 300.9 \\
 30.0  &  90.96377 & 112.9  & 113.7 & 300.0677 & 300.1 & 300.2 \\
\hline
\end{tabular}\\[2pt]
Table~1. The tree masses and the full one-loop masses of $h^0$ and $H^0$
are compared with those given by Dablestein \cite{Dab} for
the  values of the input parameters given in (1).
\end{table*}

\begin{table*}[htb]
\label{table:eq2}
\renewcommand{\tabcolsep}{1.2pc} % enlarge column spacing
\begin{tabular}{|r|r|r|r|r|r|r|}
\hline
$\tan\beta$ & \multicolumn{1}{c|}{$m_{h^0}$}
            & \multicolumn{2}{c|}{$M_{h^0}$}
            & \multicolumn{1}{c|}{$m_{H^0}$}
            & \multicolumn{2}{c|}{$M_{H^0}$} \\
\hline
            & \multicolumn{1}{c|}{tree} & 1-loop & Dabelstein
            & \multicolumn{1}{c|}{tree} & 1-loop & Dabelstein \\
\hline
  0.5  &  51.18954 &  79.0  &  79.2 & 213.7632 & 243.8 & 243.6 \\
  2.0  &  51.18954 &  69.2  &  69.5 & 213,7632 & 216.7 & 216.7 \\
  5.0  &  82.65397 &  95.5  &  95.7 & 203.6747 & 204.4 & 204.4 \\
 10.0  &  88.93069 & 101.2  & 101.5 & 201.0134 & 201.3 & 201.3 \\
 30.0  &  90.93177 & 103.6  & 104.2 & 200.1162 & 200.2 & 200.2 \\
\hline
\end{tabular}\\[2pt]
Table~2. The tree masses and the full one-loop masses of $h^0$ and $H^0$
are compared with those given by Dablestein \cite{Dab} for
the  values of the input parameters given in (2).
\end{table*}

\subsection{Application}

As an application, we consider the production of the lighter CP-even Higgs
 $h^0$
~\cite{Hol}.  First, we compare the CP-even Higgs masses, $M_{h^0}$ and
$M_{H^0}$, 
with the results of the Dabelstein \cite{Dab}.  In Table~1 and Table~2, the
tree masses, one-loop masses are compared using parameters in (1) and (2).  We
can claim the agreement is satisfactory.

\begin{eqnarray}
 M_Z=91.187, ~M_W=80.35, ~m_t=175, &\nonumber\\
 M_{A^0}=300, ~\mu=-100, ~ M_2=400, &\nonumber \\
 \tilde m_{\tilde f_L}=\tilde m_{\tilde f_R}=m_{sf}=500, ~
   \theta_f=0 &\\ 
 {\rm for~ all~ sfermions}, &\nonumber \\
 {\rm (masses~ in~ GeV)}. &\nonumber
\end{eqnarray}

\begin{equation}
 M_{A^0}=200, ~m_t=150 ~~~{\rm (in~GeV)},
\end{equation}

\begin{figure}[htb]
\vspace{9pt}
\includegraphics*[width=17pc]{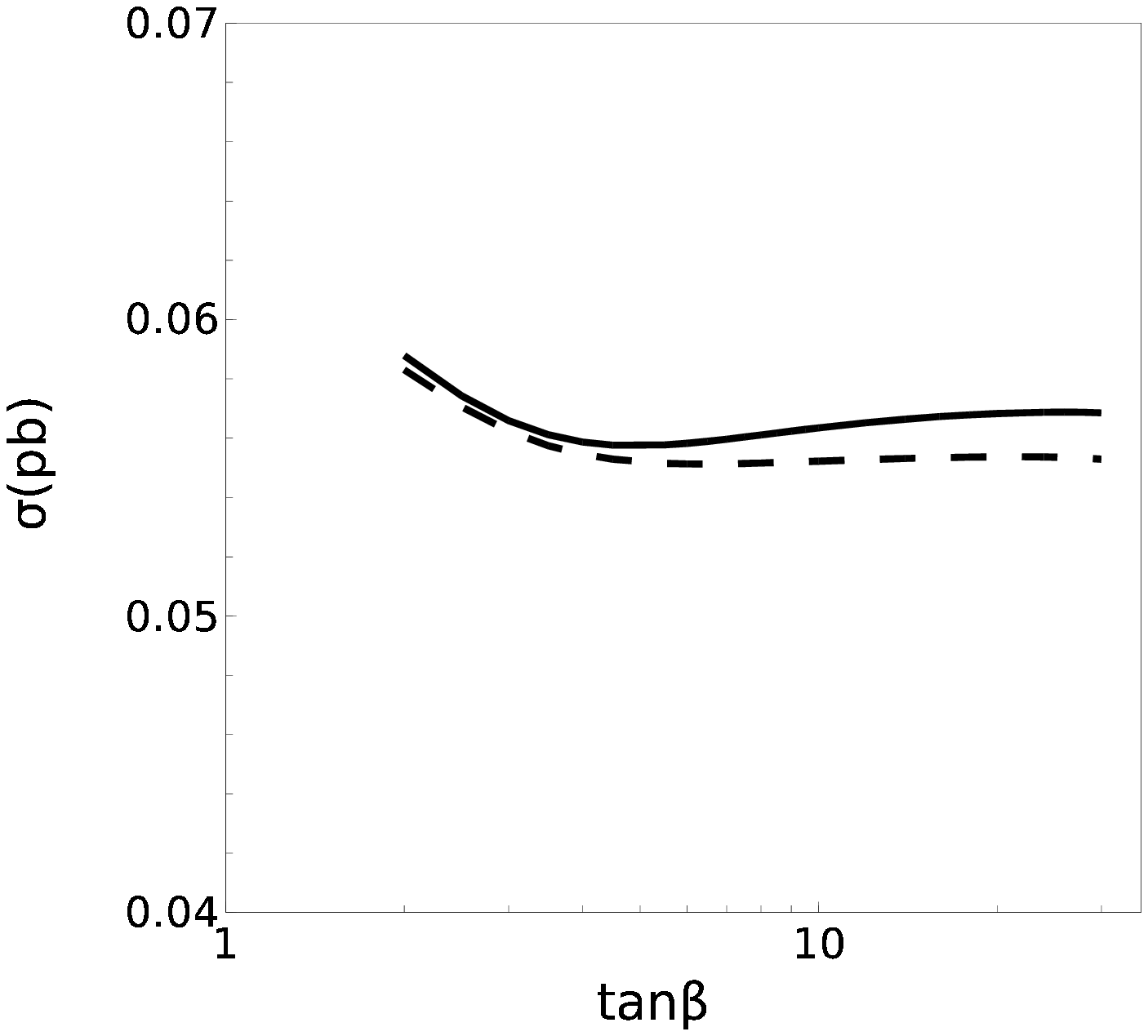}
\caption{Cross-sections for $e^+ e^- \rightarrow Z^0 h^0$~.  The solid (dashed)
line shows the radiative corrected (Born) cross sections in pb at
$\sqrt{s}=500$ GeV.}
\label{fig:z0h0}
\end{figure}

We have investigated the process $e^+ e^- \rightarrow Z^0 h^0$~\cite{Hol},
using the input values (1) except $M_W=80.423$ GeV, $m_t=174$ GeV,
~$M_{A^0}=150, ~250$ and $350$ GeV.  We found that the cross sections are not
sensitive to $M_{A^0}$ at $\sqrt{s}=500$ GeV. In Figure~\ref{fig:z0h0} we show
the results for $M_{A^0}=150$ GeV.

\section{Conclusion and outlook}
We have developed the system for the automatic computation of cross sections,
{\tt GRACE/SUSY}, including the one-loop calculations for processes in the MSSM.
For an application, we investigate the process $e^+ e^- \rightarrow Z^0 h^0$~.

Remaining tasks for us are:
\begin{itemize}
\item checking {\tt GRACE/SUSY/1LOOP} with the non-linear gauge in the MSSM
\item checking {\tt GRACE/SUSY/1LOOP} for the invariance of cross sections on
the UV constant in other processes, for example, sfermion productions and
neutralino productions.
\end{itemize}

\section*{Acknowledgements}
This work was partly supported by Japan Society for Promotion of Science
under the Grant-in-Aid for Scientific Research B ( No.14340081 ).

\end{document}